\documentclass[10pt, conference, compsocconf]{IEEEtran}
\usepackage[numbers,sort&compress]{natbib}
\usepackage{algorithm}
\usepackage{algpseudocode}
\usepackage{amsthm}
\usepackage{amsmath}
\usepackage{subfigure}
\usepackage{graphicx}
\usepackage{amsfonts}

\newtheorem{definition}{Definition}
\newtheorem{lemma}{Lemma}

\begin{document}

\title{Novel Polynomial Basis and Its Application to Reed-Solomon Erasure Codes}

\author{\IEEEauthorblockN{Sian-Jheng Lin, Wei-Ho Chung}
\IEEEauthorblockA{Research Center for Information Technology Innovation\\
 Academia Sinica\\
 Taipei City, Taiwan \\Email: sjhenglin@gmail.com; whc@citi.sinica.edu.tw}
\and
\IEEEauthorblockN{Yunghsiang S. Han}
\IEEEauthorblockA{Department of Electrical Engineering\\ National Taiwan University of Science and Technology\\ Taipei City, Taiwan\\
Email: yshan@mail.ntust.edu.tw}
}

\maketitle
\begin{abstract}
In this paper, we present a new basis of polynomial over finite fields of characteristic two and then apply it to the encoding/decoding of Reed-Solomon erasure codes. The proposed polynomial basis allows that $h$-point polynomial evaluation can be computed in $O(h\log_2(h))$ finite field operations with small leading constant. As compared with the canonical polynomial basis, the proposed basis improves the arithmetic complexity of addition, multiplication, and the determination of polynomial degree from $O(h\log_2(h)\log_2\log_2(h))$ to $O(h\log_2(h))$. Based on this basis, we then develop the encoding and erasure decoding algorithms for the $(n=2^r,k)$ Reed-Solomon codes. Thanks to the efficiency of transform based on the polynomial basis, the encoding can be completed in $O(n\log_2(k))$ finite field operations, and the erasure decoding in $O(n\log_2(n))$ finite field operations. To the best of our knowledge, this is the first approach supporting Reed-Solomon erasure codes over characteristic-2 finite fields while achieving a complexity of $O(n\log_2(n))$, in both additive and multiplicative complexities. As the complexity leading factor is small, the algorithms are advantageous in practical applications.
\end{abstract}
\thispagestyle{plain}
\pagestyle{plain}

\section{Introduction}
For a positive integer $r\geq 1$, let $\mathbb{F}_{2^r}$ denote a characteristic-2 finite field containing $2^r$ elements. A polynomial over $\mathbb{F}_{2^r}$ is defined as
\[
a(x)=a_0+a_1x+a_2x^2+\dots +a_{h-1}x^{h-1},
\]
where each $a_i\in\mathbb{F}_{2^r}$. A fundamental issue is to reduce the computational complexities of arithmetic operations over polynomials. Many fast polynomial-related algorithms, such as Reed-Solomon codes, are based on fast Fourier transforms (FFT). However, it is algorithmically harder as the traditional fast Fourier transform (FFT) cannot be applied directly over a characteristic-2 finite fields. To the best of our knowledge, no existing algorithm for characteristic-2 finite field FFT/polynomial multiplication has provably achieved $O(h\lg(h))$ operations\footnote{Throughout this paper, the notation $\lg(x)$ represents the logarithm to the base $2$.} (see Section \ref{sec:literature} for more details).

In algorithmic viewpoint, FFT is a  polynomial evaluations at a period of consecutive points, where the polynomial is in monomial basis. This viewpoint gives us the ability to design fast polynomial-related algorithms. In this paper, we present a new polynomial basis in the polynomial ring $\mathbb{F}_{2^r}[x]/(x^{2^r}-x)$. Then a transform in the new basis is defined to compute the polynomial evaluations. The new basis possesses a recursive structure which can be exploited to compute the polynomial evaluations at a period of $h$ consecutive points in time $O(h\lg(h))$ with small leading constant. Furthermore, the recursive structure also works in formal derivative with time complexity $O(h\lg(h))$.

An application of the proposed polynomial basis is in erasure codes, that is an error-correcting code by converting a message of $k$ symbols into a codeword with $n$ symbols such that the original message can be recovered from a subset of the $n$ symbols. An $(n,k)$ erasure code is called Maximum Distance Separable (MDS) if any $k$ out of the $n$ codeword symbols are sufficient to reconstruct the original message. A typical class of MDS codes is Reed-Solomon (RS) codes \cite{1960}. Nowadays, RS codes have been applied to many applications, such as RAID systems \cite{Patterson:1988,aliev2014redundant}, distributed storage codes \cite{Huang:2012,Sathiamoorthy:2013}, and data carousel \cite{Byers:1998}. Hence, the computational complexity of RS erasure code is considered crucial and has attracted substantial research attention. Based on the new polynomial basis, this paper presents the encoding/decoding algorithms for RS erasure codes. The proposed algorithms use the structure \cite{SJLin2013} that requires evaluating a polynomial and it's derivatives, while the polynomial used in the structure is in the new polynomial basis, rather than the monomial basis.

The rest of this paper is organized as follows. The proposed polynomial basis is defined in Section \ref{sec:fastTransform}. Section \ref{FastTransform} gives the definition and algorithm of the transform to compute the polynomial evaluations based on the proposed polynomial basis. Section~\ref{sec:formalderivative} shows the formal derivative of polynomial. Section \ref{sec:RSalgorithm} presents the encoding and erasure decoding algorithm for Reed-Solomon codes. The  discussions and comparisons are placed in Section \ref{Discussion}. Section\ref{sec:literature} reviews some related literature. Concluding remarks are provided in Section \ref{sec:conclusion}.

\section{A new polynomial basis over $\mathbb{F}_{2^r}$}\label{sec:fastTransform}

\subsection{Finite field arithmetic}\label{sec:Arithmetic}
Let $\mathbb{F}_{2^r}$ be an extension finite field with dimension $r$ over $\mathbb{F}_2$. The elements of $\mathbb{F}_{2^r}$ are represented as a set $\{\omega_i\}_{i=0}^{2^r-1}$. We order those elements as follows. Assume that $V$ be the $r$-dimensional vector space spanned by $v_0,v_1,\dots ,v_{r-1}\in \mathbb{F}_{2^r}$ over $\mathbb{F}_2$. For any $0\leq i<2^r$, its binary representation is given as
\begin{equation}\label{eq:binaryrepresentation}
i=i_0+i_1\cdot 2+i_2\cdot 2^2+\dots +i_{r-1}\cdot 2^{r-1}, \forall i_j\in \{0,1\}.
\end{equation}
Then  $\omega_i$ is defined as
\[
\omega_i=i_0\cdot v_0+i_1\cdot v_1+i_2\cdot v_2+\dots +i_{r-1}\cdot v_{r-1}.
\]
A polynomial $f(x)$ defined over $\mathbb{F}_{2^r}$ is a polynomial whose coefficients are from $\mathbb{F}_{2^r}$.

\subsection{Subspace vanishing polynomial}
The subspace vanishing polynomial defined in \cite{Cantor1989285,236882,5625613} is expressed as
\begin{equation}\label{eq:w_j(x)}
W_j(x)=\prod_{i=0}^{2^j-1}(x+\omega_i),
\end{equation}
where $0\leq j\leq r-1$. It can be seen that $deg(W_j(x))=2^j$.

Next we present properties of $W_j(x)$ without proof.
\begin{lemma}[\cite{236882}]\label{canonical_form_W_i(x)}
$W_j(x)$ is an $\mathbb{F}_2$-linearlized polynomial for which
\begin{equation}\label{eq:W_i(x)}
W_j(x)=\sum_{i=0}^{j}a_{j,i}x^{2^i},
\end{equation}
where each $a_{j,i}\in \mathbb{F}_{2^r}$ is a constant. Furthermore,
\begin{equation}\label{linear}
W_j(x+y)=W_j(x)+W_j(y), \forall x,y \in \mathbb{F}_{2^r}.
\end{equation}
\end{lemma}

\subsection{Polynomial basis}
In this work, we consider the polynomial ring $\mathbb{F}_{2^r}[x]/(x^{2^r}-x)$. A form of polynomial basis we work with is denoted as $\mathbb{X}(x)=\left(X_0(x),X_1(x),\dots ,X_{2^r-1}(x)\right)$ over $\mathbb{F}_{2^r}$. Each polynomial $X_i(x)$ is defined as the product of subspace vanishing polynomials. For each polynomial $X_i(x)$, $i$ is written in binary representation as
\begin{equation}\label{eq:i}
i=i_0+i_1\cdot 2+\dots +i_{r-1}\cdot 2^{r-1}, \forall i_j\in \{0,1\}.
\end{equation}
The polynomial $X_i(x)$ is then defined as
\begin{equation}\label{eq:X_i(x)}
X_i(x)=\prod_{j=0}^{r-1}\left(\frac{W_j(x)}{W_j(\omega_{2^j})}\right)^{i_j},
\end{equation}
for $0\leq i< 2^r$. Notice that $\left(\frac{W_j(x)}{W_j(\omega_{2^j})}\right)^{i_j}=1$, if $i_j=0$. It can be seen that $deg(X_i(x))=i$.

Then a form of polynomial expression $[\bullet](x)$ is given as follows.
\begin{definition}\label{polynomial-1}
A form of polynomial expression over $\mathbb{F}_{2^r}$ is defined as
\begin{equation}\label{eq:polyexp}
[D_h](x)=\sum_{i=0}^{h-1}d_iX_i(x),
\end{equation}
where
\begin{equation}\label{eq:polyvec}
D_h=(d_0, d_1,\dots , d_{h-1})
\end{equation}
is an $h$-element vector denoting the polynomial coefficients and $h\le 2^r$. Consequently, $deg([D_h](x))\leq h-1$.
\end{definition}

\section{Fast transform $\Psi_h^l[\bullet]$}\label{FastTransform}
In this section, we define a $h$-point transformation $\Psi_h^l[\bullet]$ that computes the evaluations of $[\bullet](x)$ at $h$ successive points, for $h$ a power of two. Given a $h$-element input vector $D_h$, the polynomial $[D_h](x)$ can be constructed accordingly. The transform outputs a $h$-element vector
\[
\hat{D}_h^l=\Psi_h^l[D_h],
\]
where
\[
\hat{D}_h^l=([D_h](\omega_0+\omega_l),[D_h](\omega_1+\omega_l),\dots ,[D_h](\omega_{h-1}+\omega_l)),
\]
and $l$ denotes the amount of shift in the transform.

Oppositely, the inversion, denoted as $(\Psi_h^l)^{-1}[\bullet]$, can convert $\hat{D}_h^l$ into $D_h$, and we have $(\Psi_h^l)^{-1}[\hat{D}_h^l]=D_h$. Here, we omit to provide the close form for inversion. Instead,  an algorithm for transform $\Psi_h^l[\bullet]$ and the inverse algorithm will be presented later.

\subsection{Recursive structure in polynomial basis}\label{recursive}
This subsection shows that the polynomial $[D_h](x)$ can be formulated as a recursive function $[D_h](x)=\Delta_0^0(x)$, where the function $\Delta_i^m(x)$ is defined as

\begin{equation}\label{eq:P_j^i(x)rec}
\begin{aligned}
\Delta_i^m(x)=\Delta_{i+1}^m(x)+ \frac{W_i(x)}{W_i(\omega_{2^i})}\Delta_{i+1}^{m+2^i}(x)\\
,\textup{ for }0\leq i\leq \lg(h)-1;
\end{aligned}
\end{equation}
\begin{equation}\label{eq:Delta_1,h^m(x)}
\Delta_{\lg(h)}^m(x)=d_m,\textup{ for }0\leq m\leq h-1.
\end{equation}
Note that $m$ in $\Delta_i^m(x)$ represents a $\lg(h)$-bits binary integer
\begin{equation}
m=m_0+m_1\cdot 2+\dots +m_{i-1}\cdot 2^{i}, \forall m_j\in \{0,1\}.
\end{equation}
By induction, it can be seen that $deg(\Delta_i^m(x))\leq h/2^i-1$. For example, if $h=8$, we have
{\footnotesize\begin{equation}
\begin{aligned}
&[D_8](x)=\sum_{i=0}^{7}d_iX_i(x)\\
=&d_0+ d_1\frac{W_0(x)}{W_0(\omega_1)}+ d_2\frac{W_1(x)}{W_1(\omega_2)}+ d_3\frac{W_0(x)}{W_0(\omega_1)}\frac{W_1(x)}{W_1(\omega_2)}\\
&+d_4\frac{W_2(x)}{W_2(\omega_4)}+ d_5\frac{W_0(x)}{W_0(\omega_1)}\frac{W_2(x)}{W_2(\omega_4)}+ d_6\frac{W_1(x)}{W_1(\omega_2)}\frac{W_2(x)}{W_2(\omega_4)}\\
&+d_7\frac{W_0(x)}{W_0(\omega_1)}\frac{W_1(x)}{W_1(\omega_2)}\frac{W_2(x)}{W_2(\omega_4)}\\
=&\left (d_0+ d_4\frac{W_2(x)}{W_2(\omega_4)}+ \frac{W_1(x)}{W_1(\omega_2)}\left (d_2+ d_6\frac{W_2(x)}{W_2(\omega_4)}\right )\right )\\
&+ \frac{W_0(x)}{W_0(\omega_1)}\left (d_1+ d_5\frac{W_2(x)}{W_2(\omega_4)}+ \frac{W_1(x)}{W_1(\omega_2)}\left (d_3+ d_7\frac{W_2(x)}{W_2(\omega_4)}\right )\right )\\
=&\left(\Delta_2^0(x)+\frac{W_1(x)}{W_1(\omega_2)}\Delta_2^2(x)\right)\\
&+\frac{W_0(x)}{W_0(\omega_1)}\left (\Delta_2^1(x)+\frac{W_1(x)}{W_1(\omega_2)}\Delta_2^3(x)\right)\\
=&\Delta_1^0(x)+ \frac{W_0(x)}{W_0(\omega_1)}\Delta_1^1(x)=\Delta_0^0(x).
\end{aligned}
\end{equation}}

The $\Delta_i^m(x)$ possesses the following equality that will be utilized in the algorithm:
\begin{lemma}\label{P_j^i(x)}
\begin{equation}\label{eq:Delta_j,H^i(xoplusb)}
\Delta_i^m(x+y)=\Delta_i^m(x),\forall y\in \{\omega_b\}_{b=0}^{2^i-1}.
\end{equation}
\begin{proof}
By Lemma \ref{canonical_form_W_i(x)}, we have
{\small \begin{equation}\label{eq:W_j(xoplus b)}
W_i(x+y)=W_i(x)+ W_i(y)=W_i(x),\forall y\in \{\omega_b\}_{b=0}^{2^i-1}.
\end{equation}}

The proof follows mathematical induction on $i$. In the base case, we consider \eqref{eq:P_j^i(x)rec} at $i=\lg(h)-1$:
\begin{align*}
&\Delta_{\lg(h)-1}^m(x)\\
=&\Delta_{\lg(h)}^m(x)+ \frac{W_{\lg(h)-1}(x)}{W_{\lg(h)-1}(\omega_{2^{\lg(h)-1}})}\Delta_{\lg(h)}^{m+2^{\lg(h)-1}}(x)\\
=&d_m+ \frac{W_{\lg(h)-1}(x)}{W_{\lg(h)-1}(\omega_{2^{\lg(h)-1}})}d_{m+2^{\lg(h)-1}}.
\end{align*}

From \eqref{eq:W_j(xoplus b)}, we have
\begin{align*}
&\Delta_{\lg(h)-1}^m(x+y)\\
=&d_m+ \frac{W_{\lg(h)-1}(x+y)}{W_{\lg(h)-1}(\omega_{2^{\lg(h)-1}})}d_{m+2^{\lg(h)-1}}\\
=&d_m+ \frac{W_{\lg(h)-1}(x)}{W_{\lg(h)-1}(\omega_{2^{\lg(h)-1}})}d_{m+2^{\lg(h)-1}}\\
=&\Delta_{\lg(h)-1}^m(x),\forall y\in \{\omega_b\}_{b=0}^{h/2-1}.
\end{align*}
Thus \eqref{eq:Delta_j,H^i(xoplusb)} holds for $i=\lg(h)-1$.

Assume \eqref{eq:Delta_j,H^i(xoplusb)} holds for $i=c+1$. When $i=c$, we have
\begin{align*}
&\Delta_c^m(x+y)\\
=&\Delta_{c+1}^m(x+y)+ \frac{W_c(x+y)}{W_c(\omega_{2^c})}\Delta_{c+1}^{m+2^c}(x+y)\\
=&\Delta_{c+1}^m(x+y)+ \frac{W_c(x)}{W_c(\omega_{2^c})}\Delta_{c+1}^{m+2^c}(x+y)\\
=&\Delta_{c+1}^m(x)+ \frac{W_c(x)}{W_c(\omega_{2^c})}\Delta_{c+1}^{m+2^c}(x)\\
=&\Delta_c^m(x), \forall y\in \{\omega_b\}_{b=0}^{2^c-1}.&
\end{align*}
This completes the proof.
\end{proof}
\end{lemma}
\subsection{Proposed algorithm}\label{sec:algorithm}
Let
\begin{equation}\label{eq:values2}
\begin{aligned}
\Psi(i,m,l)=\{\Delta_i^m(\omega_c+\omega_l)|c\in \{b\cdot 2^i\}_{b=0}^{h/2^i-1}\}\\
,\textup{ for }0\leq i\leq \lg(h)-1;
\end{aligned}
\end{equation}
\begin{equation}\label{eq:values3}
\Psi(\lg(h),m,l)=\{d_m\}.
\end{equation}
The objective of algorithm is to compute the values in set $\Psi(0,0,l)$. In the following, we rearrange the set $\Psi(i,m,l)$ into two parts: $\Psi(i+1,m,l)$ and $\Psi(i+1,m+2^i,l)$, by taking around $h/2^i$ additions and $h/2^{i+1}$ multiplications.

In \eqref{eq:values2}, $\Psi(i,m,l)$ can be divided into two individual subsets:
\begin{equation}\label{eq:delta0}
\{\Delta_i^m(\omega_c+\omega_l)|c\in \{b\cdot 2^{i+1}\}_{b=0}^{h/2^{i+1}-1}\}
\end{equation}
and
\begin{equation}\label{eq:delta1}
\{\Delta_i^m(\omega_c+\omega_l+\omega_{2^i})|c\in \{b\cdot 2^{i+1}\}_{b=0}^{h/2^{i+1}-1}\}.
\end{equation}
In \eqref{eq:delta0}, we have
{\begin{equation}\label{eq:Deltajimcoplusl}
\begin{aligned}
&\Delta_i^m(\omega_c+\omega_l)\\
=&\Delta_{i+1}^m(\omega_c+\omega_l)+ \frac{W_i(\omega_c+\omega_l)}{W_i(\omega_{2^i})}\Delta_{i+1}^{m+2^i}(\omega_c+\omega_l).
\end{aligned}
\end{equation}}
It can be seen that $\Delta_{i+1}^m(\omega_c+\omega_l)\in\Psi(i+1,m,l)$, and $\Delta_{i+1}^{m+2^i}(\omega_c+\omega_l)\in\Psi(i+1,m+2^i,l)$. The factor
$\frac{W_i(\omega_c+\omega_l)}{W_i(\omega_{2^i})}$
can be precomputed and stored. Hence, for each element of the set given in~\eqref{eq:delta0}, the calculation requires a multiplication and an addition. Note that when $\omega_c+\omega_l=0$, we have
\begin{equation}\label{eq:Deltajim0}
\Delta_i^m(0)=\Delta_{i+1}^m(0),
\end{equation}
which does not involve any arithmetic operations.

Next we consider the computation in \eqref{eq:delta1}, and we have
\begin{equation}\label{eq:Deltajimcoplusloplusi}
\begin{aligned}
&\Delta_i^m(\omega_c+\omega_l+\omega_{2^i})=\Delta_{i+1}^m(\omega_c+\omega_l+\omega_{2^i})\\
&+ \frac{W_i(\omega_c+\omega_l+\omega_{2^i})}{W_i(\omega_{2^i})}\Delta_{i+1}^{m+2^i}(\omega_c+\omega_l+\omega_{2^i}).
\end{aligned}
\end{equation}
By Lemma \ref{P_j^i(x)}, we have
\[
\Delta_{i+1}^m(\omega_c+\omega_l+\omega_{2^i})=\Delta_{i+1}^m(\omega_c+\omega_l);
\]
\[
\Delta_{i+1}^{m+2^i}(\omega_c+\omega_l+\omega_{2^i})=\Delta_{i+1}^{m+2^i}(\omega_c+\omega_l).
\]
Furthermore, the factor can be rewritten as
\begin{align*}
&\frac{W_i(\omega_c+\omega_l+\omega_{2^i})}{W_i(\omega_{2^i})}\\
=&\frac{W_i(\omega_c+\omega_l)+W_i(\omega_{2^i})}{W_i(\omega_{2^i})}\\
=&\frac{W_i(\omega_c+\omega_l)}{W_i(\omega_{2^i})}+1.
\end{align*}
With above results, \eqref{eq:Deltajimcoplusloplusi} can be rewritten as
\begin{equation}\label{eq:Deltajimcoplusloplusi2}
\begin{aligned}
&\Delta_i^m(\omega_c+\omega_l+\omega_{2^i})\\
=&\Delta_{i+1}^m(\omega_c+\omega_l)+\left (\frac{W_i(\omega_c+\omega_l)}{W_i(\omega_{2^i})}+1\right )\Delta_{i+1}^{m+2^i}(\omega_c+\omega_l)\\
=&\Delta_{i+1}^m(\omega_c+\omega_l)+\frac{W_i(\omega_c+\omega_l)}{W_i(\omega_{2^i})}\Delta_{i+1}^{m+2^i}(\omega_c+\omega_l)\\
&+\Delta_{i+1}^{m+2^i}(\omega_c+\omega_l)\\
=&\Delta_i^m(\omega_c+\omega_l)+\Delta_{i+1}^{m+2^i}(\omega_c+\omega_l).
\end{aligned}
\end{equation}
Hence, the element requires an addition.

\subsection{Inverse transform}
The inversion is a transform converts $\Psi(i,m,l)$ into polynomial coefficients $\{d_m\}_{m=0}^{h-1}$. The inversion can be done through backtracking the transform algorithm. As mentioned previously, $\Psi(i,m,l)$ can be rearranged into two parts: $\Psi(i+1,m,l)$ and $\Psi(i+1,m+2^i,l)$. Assume the set $\Psi(i,m,l)$ is given, we present the method to compute $\Psi(i+1,m,l)$ and $\Psi(i+1,m+2^i,l)$, respectively.

To construct $\Psi(i+1,m+2^i,l)$, \eqref{eq:Deltajimcoplusloplusi2} is reformulated as
\begin{equation}\label{eq:Deltaj22imiomegacomegal}
\Delta_{i+1}^{m+2^i}(\omega_c+\omega_l)=\Delta_i^m(\omega_c+\omega_l)+\Delta_i^m(\omega_c+\omega_l+\omega_{2^i}).
\end{equation}
Since $\Delta_i^m(\omega_c+\omega_l),\Delta_i^m(\omega_c+\omega_l+\omega_{2^i})\in\Psi(i,m,l)$, each $\Delta_{i+1}^{m+2^i}(\omega_c+\omega_l)\in\Psi(i+1,m+2^i,l)$ can be calculated with taking an addition.

To construct $\Psi(i+1,m,l)$, \eqref{eq:Deltajimcoplusl} is reformulated as
\begin{equation}\label{eq:Deltaj22imomegacomegal}
\begin{aligned}
&\Delta_{i+1}^m(\omega_c+\omega_l)\\
=&\Delta_i^m(\omega_c+\omega_l)+\frac{W_i(\omega_c+\omega_l)}{W_i(\omega_{2^i})}\Delta_{i+1}^{m+2^i}(\omega_c+\omega_l).
\end{aligned}
\end{equation}
Since $\Delta_i^m(\omega_c+\omega_l)\in\Psi(i,m,l)$ and $\Delta_{i+1}^{m+2^i}(\omega_c+\omega_l)\in\Psi(i+1,m+2^i,l)$ are known, each $\Delta_{i+1}^m(\omega_c+\omega_l)\in\Psi(i+1,m,l)$ can be calculated with taking an addition and a multiplication.

\begin{figure}
\center
\subfigure[The transform.]{
   \includegraphics[width=1.0\columnwidth]{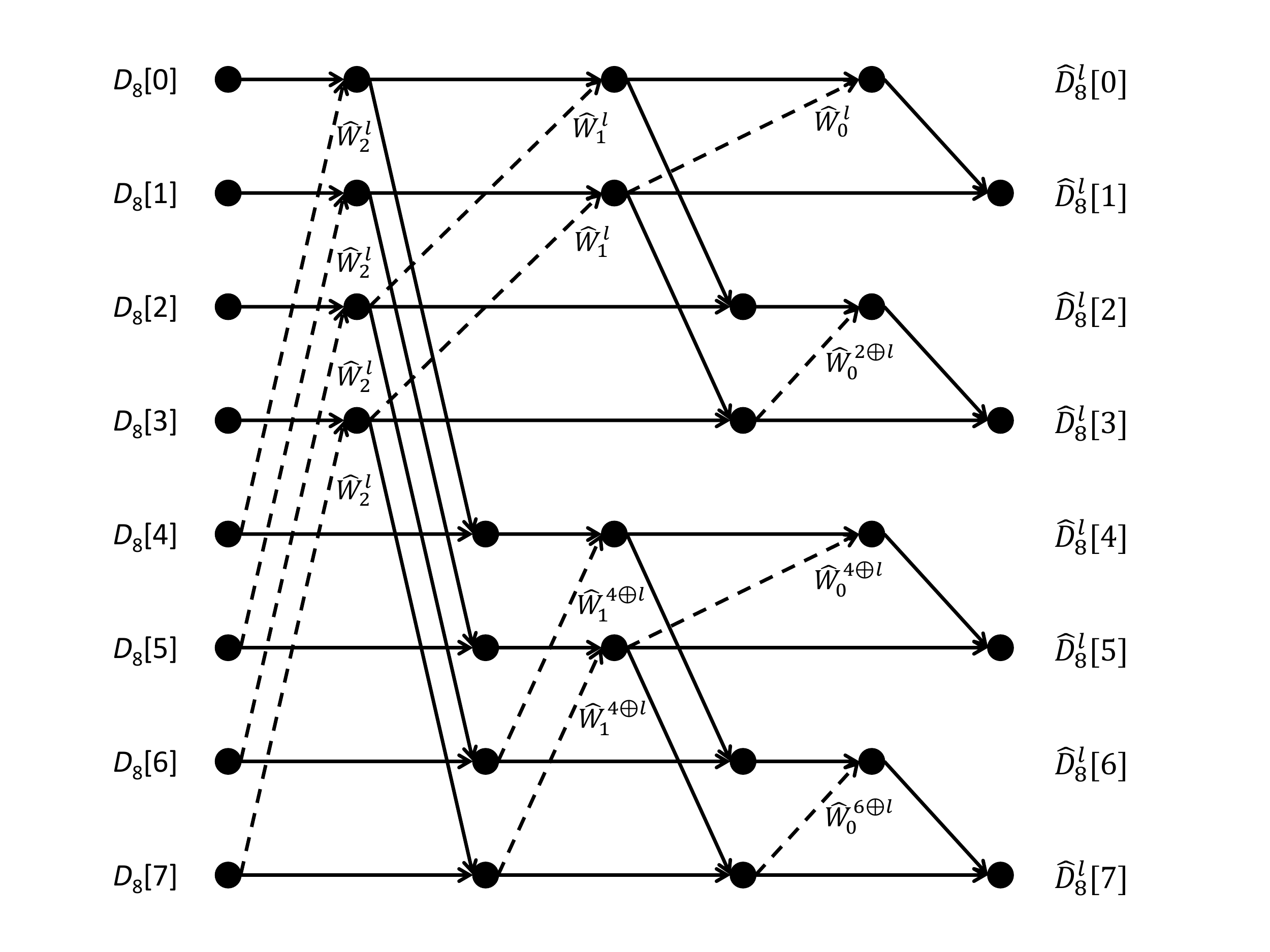}
   \label{fig:subfig1}
}
\subfigure[The inverse transform.]{
   \includegraphics[width=1.0\columnwidth]{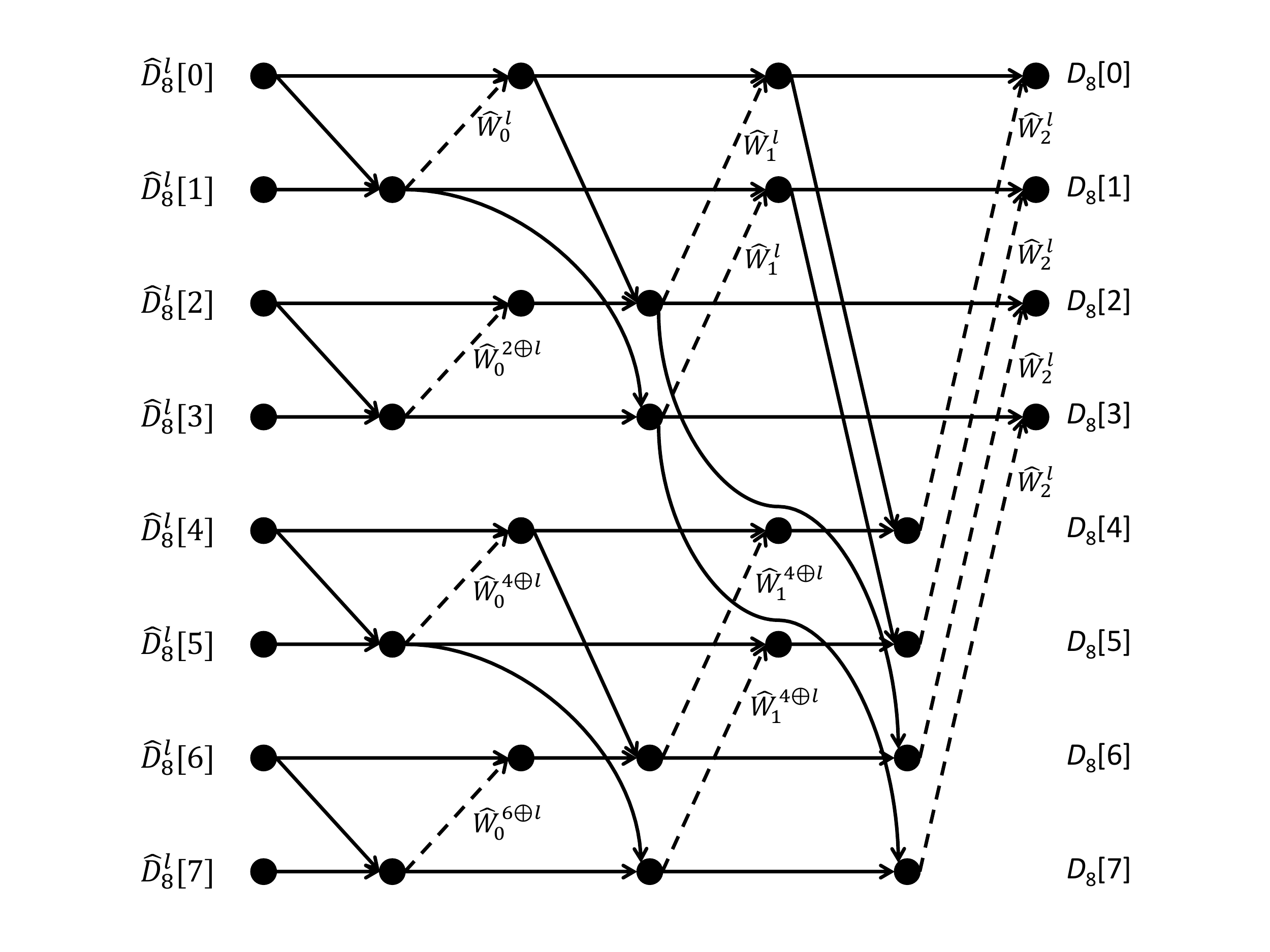}
   \label{fig:subfig2}
}
\caption{\label{fig:fig1}Data flow diagram of proposed transform of length $h=8$.}
\end{figure}

Figure \ref{fig:fig1} depicts an example of the proposed transform $\Psi_h^l[\bullet]$ of length $h=8$. Figure \ref{fig:subfig1} shows the flow graph of the transform. The dotted line arrow denotes that the element should be multiplied with a scalar factor $\hat{W}_i^j$ upon adding together with other element, where the scalar factor is denoted as
\[
\hat{W}_i^j=\frac{W_i(\omega_j)}{W_i(\omega_{2^i})}.
\]
Figure \ref{fig:subfig2} shows the flow graph of inversion. Also, it would be of interest to compare Figure \ref{fig:fig1} with the butterfly diagram of radix-2 FFT.

\subsection{Computational complexity}
Clearly, the proposed transform and its inversion have the same computational complexity. Thus, we only consider the computational complexity on transform. By the recursive structure, the number of arithmetic operations can be formulated as recursive functions. Let $A(h)$ and $M(h)$ respectively denote the number of additions and multiplications used in the algorithm. By \eqref{eq:Deltajimcoplusl} and \eqref{eq:Deltajimcoplusloplusi2}, the recursive formula is given by
\begin{align*}
A(h)=2A(h/2)+h;A(1)=0;\\
M(h)=2M(h/2)+h/2;M(1)=0.\\
\end{align*}
The solution is
\[
A(h)=h\lg{(h)};\ M(h)=\frac{h}{2}\lg{(h)}.
\]
Notice that when the amount of shift $\omega_l=0$, the number of operations can be reduced slightly (see \eqref{eq:Deltajim0}). In this case, we have
\[
A_0(h)=h\lg{(h)}-h+1;\ M_0(h)=\frac{h}{2}\lg{(h)}-h+1.
\]
\subsection{Space complexity}
In a $h$-point transform, we need $h$ units of space for the input data and an array to store the factors used in the computation of \eqref{eq:delta0}. From \eqref{eq:Deltajimcoplusl}, the factors are
\[
\frac{W_i(\omega_c+\omega_l)}{W_i(\omega_{2^i})}=\frac{W_i(\omega_c)}{W_i(\omega_{2^i})}+\frac{W_i(\omega_l)}{W_i(\omega_{2^i})}, \forall c\in \{b\cdot 2^{i+1}\}_{b=0}^{h/2^{i+1}-1}.
\]
As $0\leq i\leq\lg(h)$, a $h$-point transform requires a total of
\[
\frac{h}{2}+\frac{h}{4}+\dots +\frac{h}{h}=h-1
\]
units of space to store the factors. Hence, the space complexity is $O(h)$.

\section{Formal derivative}\label{sec:formalderivative}
In this section, we consider the formal derivative over the proposed basis. Section \ref{sec:formula} gives the closed form of the formal derivative. Section\ref{sec:computationmethod} presents a computation method that has lower multiplicative complexity than the original approach.

\subsection{Closed-form expression of formal derivative of $[D_h](x)$}\label{sec:formula}
\begin{lemma}\label{W'_j(x)}
The formal derivative of $W_i(x)$ is a constant given by
\begin{equation}
W_i'(x)=\prod_{j=1}^{2^i-1}\omega_j.
\end{equation}
\begin{proof}
Let
\[
C(x)=c\cdot x^j,
\]
where $c\in \mathbb{F}_{2^r}$. Its formal derivative is defined as
\[
C'(x)=\left\{\begin{matrix}
0 & \textrm{if $j$ is even;}\\
cx^{j-1} & \textrm{otherwise.}
\end{matrix}\right.
\]
From Lemma \ref{canonical_form_W_i(x)}, $W_i(x)$ has terms in the degrees of $1,2,4,\dots ,2^i$, so the formal derivative of $W_i(x)$ is a constant that is the coefficient of $W_i(x)$ at degree $1$. The value is
\[
\sum_{l=0}^{2^i-1}\prod_{j\neq l}\omega_j=\prod_{j=1}^{2^i-1}\omega_j.
\]
This completes the proof.
\end{proof}
\end{lemma}
By Lemma \ref{W'_j(x)}, the formal derivative of $X_i(x)$ is shown to be
\begin{equation}\label{eq:X'_i(x)}
\begin{aligned}
X_i(x)=&\sum_{l=0}^{r-1}i_l\frac{W_l'(x)}{W_l(\omega_{2^l})}\prod_{j\neq l}\left(\frac{W_j(x)}{W_j(\omega_{2^j})}\right)^{i_j}\\
=&\sum_{l\in I_i}W'_l\cdot X_{i-2^l}(x),
\end{aligned}
\end{equation}
where
\begin{equation}
W'_l=\frac{W_l'(x)}{W_l(\omega_{2^l})}=\frac{\prod_{j=1}^{2^l-1}\omega_j}{W_l(\omega_{2^l})},\label{W-l}
\end{equation}
and $I_i$ is a set including all the non-zero indices in the binary representation of $i$, given by
\[
I_i=\{j|i_j=1, j=0,1,\dots ,r-1\}.
\]
For example, if $i=13=2^0+2^2+2^3$, we have
\begin{equation}
\begin{aligned}
&X'_{13}(x)\\
=&W'_0W_2(x)W_3(x)+ W'_2W_0(x)W_3(x)+W'_3W_0(x)W_2(x)\\
=&W'_0X_{12}(x)+ W'_2X_{9}(x)+ W'_3X_{5}(x).
\end{aligned}
\end{equation}

From \eqref{eq:X'_i(x)}, the formal derivative of $[D_h](x)$ is given by
\begin{equation}\label{D'_h(x)-1}
[D_h]'(x)=\sum_{i=0}^{h-1}d_i\sum_{l\in I_i}W'_l\cdot X_{i-2^l}(x),
\end{equation}
We move the term $X_j(x)$ out of the summation operator to get
\begin{equation}\label{D'_h(x)-2}
[D_h]'(x)=\sum_{j=0}^{h-1}X_j(x)\sum_{l\in I_j^c}W'_l\cdot d_{j+2^l},
\end{equation}
where $I_j^c$ is the complement of $I_j$ defined as
\[
I_j^c=\{i\}_{i=0}^{\lg(h)-1}\setminus I_j.
\]

From \eqref{D'_h(x)-2}, when $W'_l$ given in \eqref{W-l} are pre-computed and stored, each coefficient of $X_j(x)$ requires at most $\lg(h)-1$ additions and $\lg(h)$ multiplications. Thus a native way to compute the formal derivation of $[D_h](x)$ requires $O(h\lg(h))$ operations, in both additive complexity and multiplicative complexity.

\subsection{Computation method with lower multiplicative complexity}\label{sec:computationmethod}
We present an alternative approach whose multiplicative complexity is lower than the above approach. Define
\begin{equation}\label{eq:dimathrmd}
d_i^{\mathrm{d}}=d_i\prod_{j\in I_i}W_j',
\end{equation}
for $0\leq i\leq h-1$. By substituting \eqref{eq:dimathrmd} into \eqref{D'_h(x)-2}, we have
\begin{equation}\label{eq:XDhx2}
[D_h]'(x)=\sum_{j=0}^{h-1}X_j(x)\sum_{l\in I_j^c}\frac{W_l'\cdot d_{j+2^l}^{\mathrm{d}}}{\prod_{m\in I_{j+2^l}}W_m'}.
\end{equation}
As
\[
\prod_{m\in I_{j+2^l}}W_m'=W_l'\prod_{m\in I_j}W_m',
\]
\eqref{eq:XDhx2} can be rewritten as
\begin{equation}\label{eq:XDhx3}
\begin{aligned}
\left[D_h\right]'(x)&=\sum_{j=0}^{h-1}X_j(x)\sum_{l\in I_j^c}\frac{d_{j+2^l}^{\mathrm{d}}}{\prod_{m\in I_j}W_m'}\\
&=\sum_{j=0}^{h-1}X_j(x)\frac{\sum_{l\in I_j^c}d_{j+2^l}^{\mathrm{d}}}{\prod_{m\in I_j}W_m'}.
\end{aligned}
\end{equation}

By the above formulas, the method of computing $[D_h]'(x)$ consists of two steps. In the first step, we compute \eqref{eq:dimathrmd}. Here, the set of factors
\begin{equation}\label{eq:B}
B=\{\prod_{j\in I_i}W_j'|i=0,1,\dots ,h-1\}
\end{equation}
can be pre-computed and stored, and this step only requires $h$ multiplications. In the second step, we compute the coefficients through \eqref{eq:XDhx3}. Notice that the denominator is an element of $B$. Thus, this step needs around $\frac{1}{2}h\lg(h)$ additions and $h$ multiplications.

Next we use an example to demonstrate how to obtain $[D_h]'(x)$. If $h=8$ and the set $B$ includes $8$ elements defined as
\begin{align*}
&B_0=1;B_1=W_0';B_2=W_1';B_3=W_0'W_1';\\
&B_4=W_2';B_5=W_0'W_2';B_6=W_1'W_2';B_7=W_0'W_1'W_2'.
\end{align*}
From \eqref{eq:dimathrmd}, each $d_i, 0\leq i\leq 7$ is computed via
\[
d_i^{\mathrm{d}}=d_iB_i.
\]
From \eqref{eq:XDhx3}, the formal derivative of $[D_8](x)$ is shown to be
\begin{align*}
&[D_8]'(x)\\
=&X_0(x)\frac{d_1^\mathrm{d}+ d_2^\mathrm{d}+ d_4^\mathrm{d}}{B_0}+X_1(x)\frac{d_3^\mathrm{d}+ d_5^\mathrm{d}}{B_1}+X_2(x)\frac{d_3^\mathrm{d}+ d_6^\mathrm{d}}{B_2}\\
&+X_3(x)\frac{d_7^\mathrm{d}}{B_3}+X_4(x)\frac{d_5^\mathrm{d}+ d_6^\mathrm{d}}{B_4}+X_5(x)\frac{d_7^\mathrm{d}}{B_5}+X_6(x)\frac{d_7^\mathrm{d}}{B_6}.
\end{align*}
\section{Algorithms of Reed-Solomon erasure codes}\label{sec:RSalgorithm}
Based on the new polynomial basis, this section presents the encoding and decoding algorithms for $(n,k)$ Reed-Solomon (RS) erasure codes over characteristic-2 fields. There are two major approaches on the encoding of Reed-Solomon codes, termed as polynomial evaluation approach and generator polynomial approach. In this paper, we follow the polynomial evaluation approach, which treats the codeword symbols as the evaluation values of a polynomial $F(x)$ of degree less than $k$. Let
\[
M_k=(m_0, m_1,\dots , m_{k-1})
\]
denote the vector of message, for each $m_i \in\mathbb{F}_{2^r}$. In the systematic construction, $F(x)$ is a polynomial of degree less than $k$ such that
\begin{equation}\label{eq:F(i)=m_i}
F(\omega_i)=m_i,\textup{ for }0\leq i\leq k-1.
\end{equation}
By the set of equations \eqref{eq:F(i)=m_i}, $F(x)$ can be uniquely constructed via polynomial interpolation. Then we use this $F(x)$ to calculate the codeword
\[
F_n=(F(\omega_0),F(\omega_1),\dots , F(\omega_{n-1})).
\]
In decoding, assume the received codeword has $n-k$ erasures $\{F(y): y\in E\}$, where $E$ denotes the set of evaluation points of erasures. With the $k$ un-erased symbols, $F(x)$ can be uniquely reconstructed via polynomial interpolation, and thus the erasures can be computed accordingly.

In the following, we illustrate the algorithms of encoding and erasure decoding for Reed-Solomon codes. The proposed algorithm is for $k$ a power of two, and $n=2^r$. The codes for other $k$ can be derived through code shortening strategy; i.e., appending zeros to message vector so that the length of the vector is power of two.

\subsection{Encoding algorithm}
\begin{algorithm}[t]
\caption{\label{alg:Encoding} Reed-Solomon encoding algorithm.}
\begin{algorithmic}[1]
\Require
A $k$-element message vector $M_k$ over $\mathbb{F}_{2^r}$.
\Ensure
An $n$-element systematic codeword $F_n$.
\State \label{eq:barM_2^K} $\bar{M}_k=(\Psi_k^0)^{-1}[M_k]$
\For{$i=1$ to $(n/k -1)$}
\State \label{eq:F_i=} $\bar{F}_i=\Psi_k^{i\cdot k}[\bar{M}_k]$
\EndFor
\State \Return $F_n=(M_k,\bar{F}_1,\bar{F}_2,\dots , \bar{F}_{\left\lceil n/k\right\rceil -1})$.
\end{algorithmic}
\end{algorithm}

Algorithm \ref{alg:Encoding} illustrates the pseudocode of the $(n,k)$ RS encoding algorithm. In Line 1, we compute the vector
\[
\bar{M}_k=(\bar{m}_0, \bar{m}_1,\dots , \bar{m}_{k-1}),
\]
which can be formed as a polynomial
\[
[\bar{M}_k](x)=\sum_{i=0}^{k-1}\bar{m}_iX_i(x).
\]
Since $deg([\bar{M}_k](x))\leq k-1$ and
\begin{equation}\label{eq:barM_2^K(i)}
[\bar{M}_k](\omega_i)=m_i,\textup{ for }0\leq i\leq k-1
\end{equation}
we conclude that $[\bar{M}_k](x)=F(x)$. Thus, the parity-check symbols can be computed by applying the proposed transform on $\bar{M}_k$ (see Lines 2-4). The parity-check symbols are obtained in blocks with size $k$ and there are $n/k-1$ blocks.\footnote{Since $k$ and $n$ are both powers of $2$, $n$ is divisible by $k$.} For each block, the vector $\bar{F}_i$ includes $k$ elements and each element is
\[
\bar{F}_i[j]=[\bar{M}_k](\omega_{j+ (i\cdot k)})=[\bar{M}_k](\omega_{j}+\omega_{i\cdot k}), \textup{ for }0\leq j\leq k-1.
\]
In Line 5, we assemble those vectors to get the codeword vector $F_n$.

In summary, the encoding algorithm requires a $k$-element inversion $(\Psi_k^0)^{-1}[\bullet]$ and $( n/k  -1)$ times of $k$-element transform $\Psi_k^i[\bullet]$. Thus, the encoding algorithm has the complexity
\[
O((n/k)k\lg{(k)})=O(n\lg{(k)}).
\]
\subsection{Erasure decoding algorithm}
\begin{algorithm}[t]
\caption{\label{alg:Decoding} Framework of Reed-Solomon erasure decoding algorithm.}
\begin{algorithmic}[1]
\Require
Received codeword $\bar{F}_n$, and the positions of erasures $E=\{e_i\}_{i=0}^{n-k-1}$.
\Ensure
The erasures $\{F(j)|j\in E\}$.

\State Compute two sets of values $\bar{\Pi}$ and $\Pi'$, defined in \eqref{eq:Pi(j)} and \eqref{eq:Pi'(j)}.
\State From \eqref{eq:hatF(l)}, compute
\[
\Phi=(\hat{F}(\omega_0), \hat{F}(\omega_1),\dots , \hat{F}(\omega_{n-1})).
\]
\State Apply $n$-point fast inverse transform on $\Phi$ to get
\[
\bar{\Phi}_n=(\Psi_n^0)^{-1}[\Phi].
\]
\State Compute the formal derivative of $\bar{\Phi}_n$. The result is denoted as $\bar{\Phi}^\mathrm{d}_n$.
\State Apply $n$-point fast transform on $\bar{\Phi}^\mathrm{d}_n$ to get
\[
\Phi^\mathrm{d}_n=\Psi_n^0[\bar{\Phi}^\mathrm{d}_n].
\]
\State Compute the erasures via
\[
F(j)=\frac{\Phi^\mathrm{d}_n[j]}{\Pi'(j)}, \forall j\in E.
\]
\end{algorithmic}
\end{algorithm}

The decoding algorithm follows our previous work \cite{SJLin2013} that requires evaluating a polynomial and it's derivatives. The code proposed in \cite{SJLin2013} is based on Fermat number transforms (FNT). In this paper, we replace the role of FNT over $\mathbb{F}_{2^r+1}$ with the proposed transform over $\mathbb{F}_{2^r}$. However, since the proposed transform is not Fourier transform, some arithmetic operations involved in the transform should be modified accordingly.

Assume the received codeword $\bar{F}_n$ has $n-k$ erasures. The set of evaluation points of erasures are denoted as
\[
E=\{\omega_{e_i}\}_{i=0}^{n-k-1}.
\]
Let
\[
\Pi(x)=\prod_{y\in E}(x+ y)
\]
denote the error locator polynomial having zeros at all erased symbols. It can be seen that $\Pi(j)=0,\forall j\in E$. Define
\[
\hat{F}(x)=F(x)\Pi(x),
\]
and the polynomial degree is $deg(\hat{F}(x))=deg(F(x))+deg(\Pi(x))\leq n-1$. The formal derivative of $\hat{F}(x)$ is
\begin{equation}\label{eq:hatF'(x)}
\hat{F}'(x)=F'(x)\Pi(x)+F(x)\Pi'(x).
\end{equation}
By substituting $x=j\in E$ into \eqref{eq:hatF'(x)}, we have
\[
\hat{F}'(j)=F(j)\Pi'(j), \forall j\in E.
\]
Hence the erasures can be computed by
\begin{equation}\label{eq:F(j)}
F(j)=\frac{\hat{F}'(j)}{\Pi'(j)}, \forall j\in E.
\end{equation}
Based on above formulas, the decoding procedure consists of three major stages: First, compute the coefficients of $\hat{F}(x)$; second, compute the formal derivative of $\hat{F}(x)$; and third, compute the erasures by \eqref{eq:F(j)}. The details are elaborated as follows.

In the first stage, we need to compute the coefficients of $\hat{F}(x)$. It can be shown that
\begin{equation}\label{eq:hatF(l)}
\hat{F}(j)=F(j)\Pi(j)=\left\{\begin{matrix}
0 & \forall j\in E;\\
F(j)\Pi(j) & \mathrm{otherwise.}
\end{matrix}\right.
\end{equation}
Here, we define
\begin{equation}\label{eq:Pi(j)}
\bar{\Pi}=\{\Pi(j)|j\in \mathbb{F}_{2^r}\backslash E\}.
\end{equation}
Appendix shows the algorithm of computing $\bar{\Pi}$ proposed by \cite{Didier2009}. Since $F(j),\ j\in \mathbb{F}_{2^r}\backslash E$ are elements of the received vector, the result of \eqref{eq:hatF(l)} can computed with $n$ multiplications after $\bar{\Pi}$ is obtained and is denoted as a vector
\[
\Phi=(\hat{F}(\omega_0), \hat{F}(\omega_1),\dots , \hat{F}(\omega_{n-1})).
\]
Then we compute
\begin{equation}\label{eq:barPhin}
\bar{\Phi}_n=(\Psi_n^0)^{-1}[\Phi].
\end{equation}
Here, the resulting vector $\bar{\Phi}_n=(\bar{\phi}_0, \bar{\phi}_1,\dots ,\bar{\phi}_{n-1})$ can be formed as a polynomial
\[
[\bar{\Phi}_n](x)=\sum_{i=0}^{n-1}\bar{\phi}_iX_i(x),
\]
where $[\bar{\Phi}_n](\omega_j)=\hat{F}(\omega_j)$, for $0\leq j\leq n-1$. That is, $[\bar{\Phi}_n](\omega_j)-\hat{F}(\omega_j)=0$, for $0\leq j\leq n-1$. Since the degree of  $[\bar{\Phi}_n](x)-\hat{F}(x)$ is at most $n-1$, it must be the zero polynomial when it has $n$ roots. Hence, we conclude $[\bar{\Phi}_n](x)=\hat{F}(x)$.

The second stage is to compute the formal derivative of $\hat{F}(x)$. Since $[\bar{\Phi}_n](x)$ is under the polynomial basis given by Definition \ref{polynomial-1}, we compute the formal derivative of $[\bar{\Phi}_n](x)$ by the method presented in Section \ref{sec:formalderivative}. Then we can obtain the result vector $\bar{\Phi}^\mathrm{d}_n=(\bar{\phi}^\mathrm{d}_0, \bar{\phi}^\mathrm{d}_1,\dots ,\bar{\phi}^\mathrm{d}_{n-1})$, and the polynomial
\[
[\bar{\Phi}^\mathrm{d}_n](x)=\sum_{i=0}^{n-1}\bar{\phi}_i^\mathrm{d}X_i(x)
\]
is the formal derivative of $[\bar{\Phi}_n](x)$.

In the final stage, we need to compute the erasures via \eqref{eq:F(j)}. Here, the denominators in \eqref{eq:F(j)} are defined as a set
\begin{equation}\label{eq:Pi'(j)}
\Pi'=\{\Pi'(j)|j\in E\},
\end{equation}
which can be constructed by the algorithm introduced in Appendix. We then compute
\begin{equation}\label{eq:Phimathrmdn}
\Phi^\mathrm{d}_n=\Psi_n^0[\bar{\Phi}^\mathrm{d}_n],
\end{equation}
where the resulting vector includes the evaluations of $\hat{F}'(j)$ for $j\in \mathbb{F}_{2^r}$; i.e., the $\Phi^\mathrm{d}_n$ is denoted as
\[
\Phi^\mathrm{d}_n=(\hat{F}'(\omega_0),\hat{F}'(\omega_1),\dots ,\hat{F}'(\omega_{n-1})).
\]
Then the erasures can be computed through \eqref{eq:F(j)}.

The decoding procedure is summarized in Algorithm \ref{alg:Decoding}. The complexity of this algorithm is dominated by Steps 1, 3, 4 and 5, whereas Steps 2 and 6 only require $O(n)$ multiplications. By the proposed fast transform algorithm, Steps 3 and 5 can be done with $O(n\lg{(n)})$ additions and $O(n\lg{(n)})$ multiplications. By the method in Section \ref{sec:formalderivative}, Step 4 requires $O(n\lg(n))$ additions and $O(n)$ multiplications. In Step 1, we use the algorithm shown in Appendix, and it can be done with $O(n\lg(n))$ modulus operations. In summary, the proposed decoding algorithm has the complexity of order $O(n\lg{(n)})$.

\section{Discussions and comparisons}\label{Discussion}
\subsection{Complexities of operations in polynomial basis}
We consider some polynomial operations in this section.
Table \ref{Complexitiesofoperations} tabulates the complexities of some polynomial operations in the monomial basis and the proposed basis over characteristic-2 finite fields. In particular, the polynomial addition is simple by adding the coefficients of two polynomials. Hence, the complexity is $O(h)$ in both basis. For the polynomial multiplication, an algorithm with order $O(h\lg(h)\lg\lg(h))$ is proposed by \cite{Schonhage1977}, in 1977. To compute $[A_h](x)\times [B_h](x)$ in the proposed basis, the result polynomial is computed via
\[
(\Psi_{2h}^l)^{-1}[\Psi_{2h}^l[A_{2h}]\star\Psi_{2h}^l[B_{2h}]],
\]
where $A_{2h}$( and $B_{2h}$) is $2h$-point vector by appending zeros to $A_{h}$( and $B_{h}$), and $\star$ denotes the operation of pairwise multiplication. Hence, the complexity is $O(h\lg(h))$.

To determine the degree polynomial in proposed basis, we scan the coefficients of $[D_h](x)$ to determine the highest degree term $d_jX_j(x), d_j\neq 0$, and thus the complexity is $O(h\lg(h))$; and so does the polynomial in monomial basis.

The formal derivative in proposed basis requires $O(h\lg(h))$ field operations shown in Section \ref{sec:formalderivative}. In contrast, the formal derivative in monomial basis only requires $O(h)$ operations.

\begin{table}
\caption{\label{Complexitiesofoperations}Complexities of operations in polynomial basis over characteristic-2 finite fields}
\centering
\begin{tabular}{|c|c|c|}
\hline
Operations & Monomial basis & Proposed basis \tabularnewline
\hline
\hline
Addition & $O(h)$ & $O(h)$\tabularnewline
\hline
Multiplication & $O(h\lg(h)\lg\lg(h))$ & $O(h\lg(h))$\tabularnewline
\hline
Polynomial degree & $O(h)$ & $O(h)$\tabularnewline
\hline
Formal derivative & $O(h)$ & $O(h\lg(h))$\tabularnewline
\hline
\end{tabular}
\end{table}
\subsection{Comparisons with Didier's approach}
In 2009, Didier \cite{Didier2009} present an erasure decoding algorithm for Reed-Solomon codes based on fast Walsh-Hadamard transforms. The algorithm \cite{Didier2009} consists of two major parts: the first part is to compute the polynomial evaluations of error locator polynomial, and the second part is to decompose the Lagrange polynomial into several logical convolutions, which are then respectively computed with fast Walsh-Hadamard transforms. The first part requires $O(n\lg(n))$ time, and the second part requires $O(n\lg^2(n))$ time, so the complexity \cite{Didier2009} is $O(n\lg^2(n))$.
In contrast, the proposed approach employs the first part in \cite{Didier2009}; in the second part, we design another decoding structure based on the proposed basis. The proposed transform only requires $O(n\lg(n))$ time, so that the proposed approach can reduce the complexity from $O(n\lg^2(n))$ to $O(n\lg(n))$.

We also implement the proposed algorithm in C and run the program on Intel core i7-950 CPU. While $n=2^{16}$, $k/n=1/2$, the program took about $1.12$ seconds to generate a codeword, and $3.06$ seconds to decode an erased codeword on average. On the other hand, we also ran the program \cite{Didier2009} written by the author on the same platform. In our simulation, the program \cite{Didier2009} took about $52.91$ seconds in both encoding and erasure decoding under the same parameter configuration. Thus, the proposed erasure decoding is around $17$ times faster than \cite{Didier2009}, while $n=2^{16}$.

\section{Literature review}\label{sec:literature}
 In the original view of \cite{1960}, the codeword of the RS code is a sequence of evaluation values of a polynomial interpreted by message. By this viewpoint, the encoding process can be treated as an oversampling process through discrete Fourier transform (DFT) over finite fields. Some studies \cite{1055516,blahut1983theory,macwilliams1977theory} indicate that, if a $O(n\lg (n))$ finite field FFT is available, the error-correction decoding can be reduced to $O(n\lg^2(n))$. An $n$-point radix-2 FFT butterfly diagram requires $n\lg(n)$ additions and $\frac{n}{2}\lg(n)$ multiplications. This FFT butterfly diagram can be directly applied on Fermat prime fields $\mathbb{F}_{2^r+1}, r\in \{1,2,4,8,16\}$. In this case, the transform, referred to as Fermat number transform (FNT), requires $n\lg(n)$ finite field additions and $\frac{n}{2}\lg(n)$ finite field multiplications. By employing FNT, a number of fast approaches \cite{1976,1055816,1055516} had been presented to reduce the complexity of encoding and decoding of RS codes. Some FNT-based RS erasure decoding algorithms had been proposed \cite{Sora20103,SJLin2013,6362135} in $O(n\lg(n))$ finite field operations. Thus far, no existing algorithm for $(n,k)$ RS codes has decoding complexity achieving lower than $\Omega(n\lg(n))$ operations, in a context of a fixed coding rate $k/n$. However, the major drawback of FNT is that it needs more space to store one extra symbol in practical implementation such that the FNT-based codes are impractical in general applications.

On the other hand, FFTs over characteristic-2 finite fields require higher complexities than $O(n\lg(n))$. Table \ref{characteristic-2 finite field FFT} tabulates the arithmetic complexities of FFT algorithms over characteristic-2 finite fields. As shown in Table \ref{characteristic-2 finite field FFT}, Gao and Mateer \cite{5625613} gave two versions of additive FFTs over $\mathbb{F}_{2^r}$ that are most likely the most efficient FFTs by far. The first is for arbitrary $r$, and the second is for $r$ a power of two. Notably, Wu's approach \cite{6097072} has very low multiplicative complexity $O(n\lg^{\lg(3/2)}(n))$, but the additive complexity is higher with complexity $O(n^2/\lg^{\lg(8/3)}(n))$. This implies that when the polynomial representation in RS codes are in monomial basis, the complexity will fail to reach $O(n\lg(n))$.

There exist faster encoding and erasure decoding approaches in some non-MDS codes. Such codes, termed as fountain codes \cite{Byers:1998}, require a little more than $k$ codeword symbols to recover the original message. Two famous classes of fountain codes are LT code \cite{1181950} and Raptor code \cite{1638543}. Due to the low complexity, fountain codes have significant merits in many applications. However, because of the randomly generated generator matrices, the hardware parallelization of fountain code is not trivial.

\begin{table*}
\caption{\label{characteristic-2 finite field FFT}Complexities of $n$-point FFT algorithms over $\mathbb{F}_{2^r}$, where $n=2^{r}-1$}
\centering
\begin{tabular}{|c|c|c|c|}
\hline
Algorithm & Restriction & Additive complexity & Multiplicative complexity\tabularnewline
\hline
\hline
Gao \cite{5625613} & $r$ is a power of two & $O(n\lg(n)\lg\lg(n))$ & $O(n\lg(n))$\tabularnewline
\hline
Cantor \cite{Cantor1989285} & $r$ is a power of two & $O(n\lg^{\lg(3)}(n))$ & $O(n\lg(n))$\tabularnewline
\hline
Gao \cite{5625613} &  & $O(n\lg^2(n))$ & $O(n\lg(n))$\tabularnewline
\hline
Wang \cite{1926}, Gathen \cite{236882} &  & $O(n\lg^2(n))$ & $O(n\lg^2(n))$\tabularnewline
\hline
Pollard \cite{1971Pollard} & $r$ is even& $O(n^{3/2})$ & $O(n^{3/2})$\tabularnewline
\hline
Wu \cite{6097072} &  & $O(n^2/\lg^{\lg(8/3)}(n))$ & $O(n\lg^{\lg(3/2)}(n))$\tabularnewline
\hline
Sarwate \cite{1675088} &  & $O(n^2)$ & $O(n\lg(n))$\tabularnewline
\hline
Naive approach &  & $O(n^{2})$ & $O(n^{2})$\tabularnewline
\hline
\end{tabular}
\end{table*}

\section{Concluding remarks}  \label{sec:conclusion}
Based on the proposed polynomial basis, we can compute the polynomial evaluations in the complexity of order $O(h\lg(h))$ with a small leading constant. This enables our capability to encode/erasure decode $(n,k)$ Reed-Solomon codes over characteristic-2 finite field in $O(n\lg(n))$ time. As the complexity leading factor is small, the algorithms are advantageous in practical applications. To the best of our knowledge, this is the first approach supporting Reed-Solomon erasure codes on characteristic-2 finite fields to achieve complexity of $O(n\lg(n))$. In addition, all the transforms employed in the Reed-Solomon algorithms can be easily implemented in parallel processing. Hence, the proposed algorithms substantially facilitate practical applications. While this paper has demonstrated the polynomial basis and operations over characteristic-2 finite fields, it is of interest to consider the case over fields with arbitrary characteristics.

\section*{Acknowledgment}
This work was supported in part by Ministry of Science and Technology of Taiwan, under grants NSC 101-2221-E-011-069-MY3, NSC 102-2221-E-001-006-MY2, and MOST 103-3113-E-110-002.

\appendix
In \cite{Didier2009}, Didier present an efficient algorithm to compute the elements in two sets \eqref{eq:Pi(j)} and \eqref{eq:Pi'(j)}. The method is presented here for the purpose of completeness. Consider the construction of $\Pi'$. The formal derivative of $\Pi(x)$ is given by
\[
\Pi'(x)=\sum_{j\in E}\prod_{y\in E, y\neq j}(x+y).
\]
By substituting $x=j\in E$ into $\Pi'(x)$, we have
\begin{equation}\label{eq:Pi'(j)2}
\Pi'(j)=\prod_{y\in E, y\neq j}(j+y)=\prod_{y\in\mathbb{F}_{2^r}, y\neq j}(j+ y)^{R(y)},
\end{equation}
where $R(x)$ is a function defined as
\begin{equation}
R(y)=\left\{\begin{matrix}
1 & \textrm{if $y\in E$;}\\
0 &  \textrm{otherwise.}
\end{matrix}\right.
\end{equation}
Define $Log(x)$ as the discrete logarithm function. For each $i\in \mathbb{F}_{2^r}^*$, we denote $Log(i)=j$ iff $i=\alpha^j$, where $\alpha$ is the primitive element of $\mathbb{F}_{2^r}^*$. Then \eqref{eq:Pi'(j)2} can be reformulated as
\[
Log(\Pi'(j))=\biguplus_{y\in\mathbb{F}_{2^r}, y\neq j}R(y)Log(j+y), \forall j\in E.
\]
Note that the symbol $\biguplus$ means the summation with normal additions. By setting $Log(0)=0$, the above equation can be rewritten as
\begin{equation}\label{eq:Log(Pi'(j))}
Log(\Pi'(j))=\biguplus_{y\in\mathbb{F}_{2^r}}R(y)Log(j+y), \forall j\in E.
\end{equation}
Upon describing the algorithm to compute \eqref{eq:Log(Pi'(j))}, we consider the construction of another set $\Pi$. In the same way, the elements of $\Pi$ can be formulated as
\begin{equation}\label{eq:Log(Pi(j))}
Log(\Pi(j))=\biguplus_{y\in\mathbb{F}_{2^r}}R(y)Log(j+y), \forall j\in \mathbb{F}_{2^r} \setminus E.
\end{equation}
With combining \eqref{eq:Log(Pi'(j))} and \eqref{eq:Log(Pi(j))}, the objective of algorithm is to compute
\begin{equation}\label{eq:Log(Pi(j))2}
Log(\Pi(j))=\biguplus_{y\in\mathbb{F}_{2^r}}R(y)Log(j+y), \forall j\in \mathbb{F}_{2^r}.
\end{equation}
In \eqref{eq:Log(Pi(j))2}, the operation $+$ is the $\mathbb{F}_{2^r}$ addition, that can be treated as exclusive-or operation. Hence, \eqref{eq:Log(Pi(j))2} is namely the logical convolution \cite{4090612}\cite{1162394}, that can be efficiently computed with fast Walsh-Hadamard transform \cite{1674569}. The algorithm is elaborated as follows.

Let $FWT_h[\bullet]$ denote the $h$-point fast Walsh-Hadamard transform (FWHT). A $h$-point FWHT requires $h\lg(h)$ additions. Define
\[
R_{2^r}=(R(0),R(1),\dots ,R(2^r-1)),
\]
\[
L_{2^r}=(0,Log(\omega_1),Log(\omega_2),\dots ,Log(\omega_{2^r-1})).
\]
The result of \eqref{eq:Log(Pi(j))2} is computed by
\begin{equation}\label{eq:Rnmathrmw}
R_{2^r}^{\mathrm{w}}=\mathrm{FWHT}_{2^r}[\mathrm{FWHT}_{2^r}[R_{2^r}]\star \mathrm{FWHT}_{2^r}[L_{2^r}]],
\end{equation}
where the operation $\star$ denotes pairwise multiplication. To further reduce the complexity, the $\mathrm{FWHT}_{2^r}[L_{2^r}]$ can be pre-computed and stored, and thus \eqref{eq:Rnmathrmw} can be done with performing two fast Walsh transforms of length $2^r$. We remark that all the above computation can be performed over modulo $2^r-1$.
After obtaining $R_{2^r}^{\mathrm{w}}$, we compute the exponent for each element of $R_{2^r}^{\mathrm{w}}$, and this step can be done via table lookup. In summary, the algorithm requires $O(2^r\lg(2^r))$ modulus additions, $O(2^r)$ modulus multiplications, and $O(2^r)$ exponentiations.
\bibliographystyle{IEEEtran}
\bibliography{IEEEabrv,refs}
\end{document}